# A LED-based Functional Light Source for the Characterization of Thin Film Solar Cells

Md. Saidul Islam[*†], Syed Farid Uddin Farhad[*†], Md. Saidul Islam[§†], Nazmul Islam Tanvir[*†], and Suravi Islam[*†]

[*]Device Design and Simulation Research, Industrial Physics Division, BCSIR Labs, Dhaka 1205, Bangladesh
[§]Institute of Fuel Research and Development (IFRD), BCSIR, Dhaka 1205, Bangladesh
[†]Bangladesh Council of Scientific and Industrial Research (BCSIR), Dhak 1205, Bangladesh

Corresponding email: sf1878@my.bristol.ac.uk, s.f.u.farhad@bcsir.gov.bd

*Abstract*—A light source of selective functionalities of wavelengths, illumination periods, and intensities is desirable for investigating performance parameters as well as the quality of different layers and interfaces of solar cells. Conventional light sources used for these types of research are expensive, space-consuming, cumbersome to work with, and have limited functionalities. We have developed a light source with variable wavelength, intensity, and illumination period to address these issues using an illumination period control unit, voltage regulator, neutral density filter, alterable light emitting diodes, etc. As a proof-of-concept, we employed our constructed light source to investigate the intensity, wavelength, illumination period modulated photovoltaic, and impedance properties of inorganic thin film solar cells such as cadmium telluride (CdTe) and copper zinc tin sulfide (CZTS) using lights of wavelength 410, 520, and 635 nm. We hope to use this light source for photophysical and photochemical studies of metal oxide materials used for renewable energy research.

## I. INTRODUCTION

A light source contains a significant value in scientific research. Variation in different parameters of light is used in different experiments. Light sources with selectable wavelengths, intensities, and illumination periods have been utilized in photocatalytic [1, 2], Photoelectrochemical [3, 4], Solar cells [5, 6], and light-responsive other optoelectronic devices [7, 8, 9, 10] diagnosis and performance studies over the years. In short, a light source is highly desirable where any photophysical or photochemical property of materials is to be investigated. Conventional light sources used for these types of research have some limitations. For example, a broadband Xenon arc lamp is costly, requires cumbersome operation and maintenance, needs an expensive filter, cannot produce variable intensity light or periodic illumination, is heavy to work with, and is so challenging to change the orientation of illumination on different type samples. To resolve these technical problems, scientists have been trying to develop light sources using different ideas and techniques. For instance, light emitting diodes (LEDs) have been used in recent times for developing light sources for photodynamic therapy [11], communication systems [12], spectral imaging [13], etc. But these light sources are not suitable for characterizing solar cells made up of different layer materials having different bandgap energies. To conduct such studies on thin film solar cells, a light source should have a selectable range of wavelengths, regulation of intensity, and adjustable illumination periods. It would be more advantageous if the light source is portable, consumes low power, flexible at changing the direction of illumination on the sample under study, and, most importantly, can be used in any research where illumination for a specific period is required. In this paper, a facile fabrication of such a light source has been proposed and realized, which is expected to advance our research works aligned with sustainable technologies for Industry 4.0 adaptation.

## II. DESIGN AND CONCEPT

The variable wavelength, intensity, and illumination period light source have been developed using an illumination period control unit (IPCU), LEDs, voltage regulator, neutral density (ND) filter, LED display, DC motor, etc. The instrument has electrical and mechanical periodic light illumination "ON/OFF" functionalities. Users can select desired illumination period in seconds or minutes with any of these functions. The IPCU provides the electrical switching function. This unit has an AVR Microcontroller (ATMega28P) and a set of timers with 5 buttons to adjust the light pulse (periodic on/off). The Microcontroller has been programmed with Flowcode version 9.0 [14]. The optional DC motor is connected to a speed controller for precise operation. An optical chopper associated with this motor provides the mechanical switching of the light illumination. This method eliminates the latency of the LED during fast switching. The alterable LEDs that provide desired wavelength of light are held inside a metallic barrel coupled with a focusing lens. In contrast, conventional Xenon arc lamps, which mimic natural light, require filtering unwanted wavelengths. They need extra equipment to produce desired light source functionalities. The light source we developed works on a DC power supply. An LED driver is integrated here to maintain a constant voltage to drive the LED. A knob regulates this voltage, which controls the LED input current to adjust light intensity. Fig. 1 shows the control units for controlling intensity and illumination period. The control units have been designed using ExpressPCB [15]. An LED display shows the input voltage level of the mounted LED having desired wavelength. A removable ND filter associated with the instrument is an add-on facility that can also be used to control the light intensity on the sample. The whole equipment is enclosed in a plastic box apart from the optional DC motor and the ND filter. The ND filter holder and its support base are also made of plastic materials. The enclosure box has the following dimension: length (L): 10.795 cm, width (W): 6.35 cm, and height (H): 4.445 cm. The filter holder's dimension is length (L): 7.62 cm, width (W): 5.08 cm, and thickness (T): 0.79375 cm. The support base that attaches the filter holder to the enclosure box has a dimension of H×W×L: 2.2225 cm×3.4925 cm×5.08 cm. The diameter of the filter lens is 4.445 cm. The LED-containing barrel is faced outside the box through a hole with a diameter of 2.54 cm. The architecture of the light source is shown in Fig. 2.

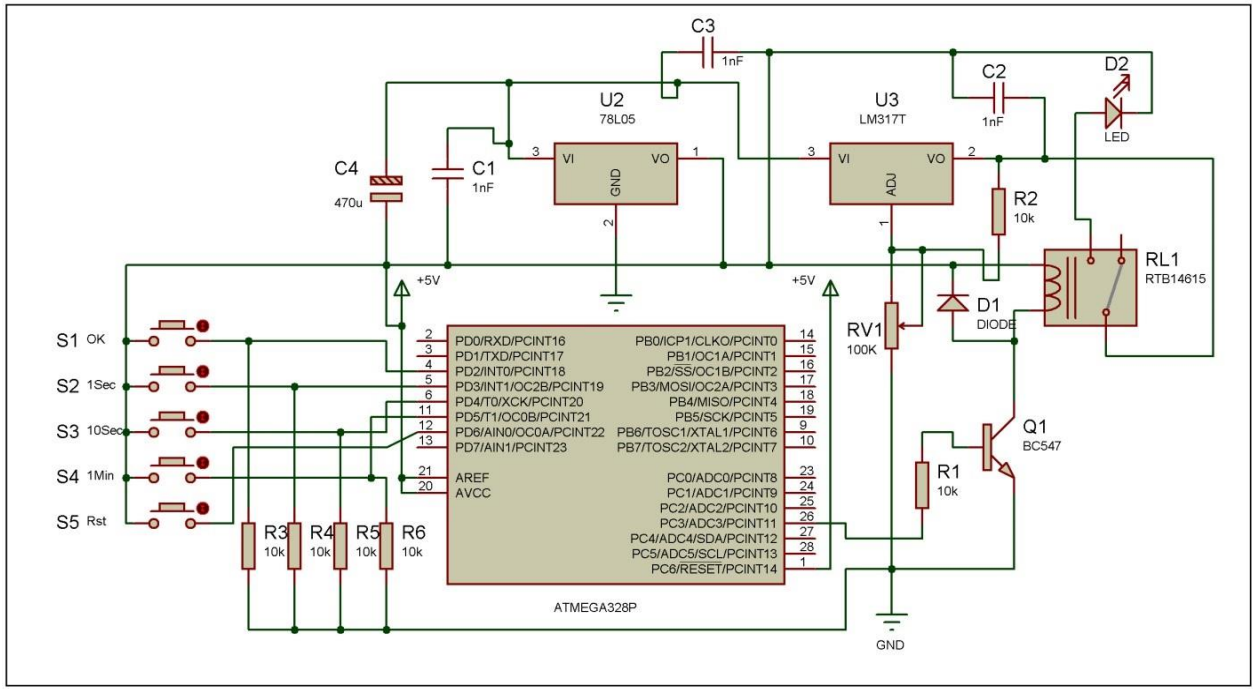

Fig. 1. Schematic diagram of the control unit for controlling intensity and illumination period of the LED-based light source.

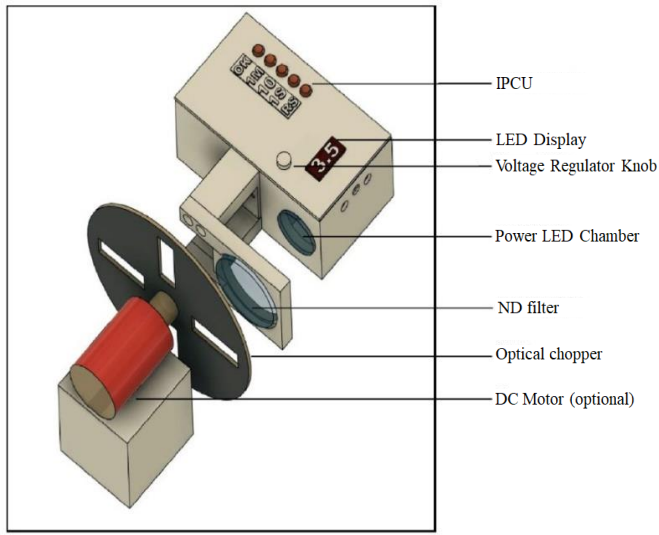

Fig. 2. Architecture of the light source.

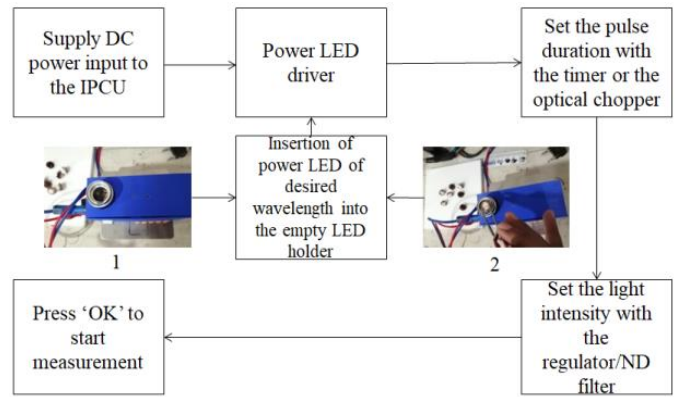

Fig. 3. Operational flowchart of the light source.

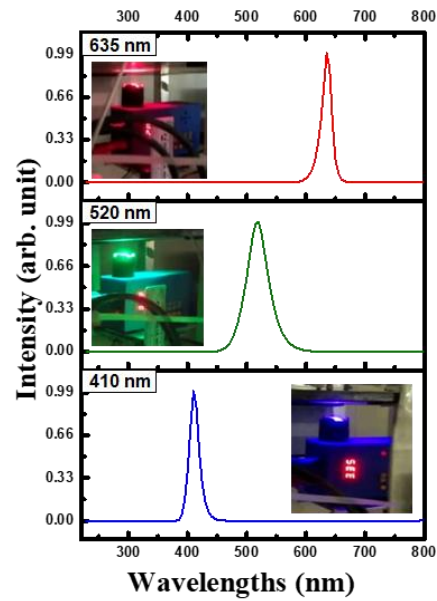

Fig. 4. Emission spectra of 410, 520, and 635 nm LEDs.

### III. WORKING PRINCIPLE

The operation of our light source is quite simple. One has to ensure the connection of the power supply, mount the required LED of the desired wavelength, set the illumination period, and regulate intensity if necessary. Fig. 3 demonstrates the flowchart of the whole operation.

LEDs of different wavelengths can be used as an illumination source. Fig. 4 demonstrates the irradiance spectra of 410, 520, and 635 nm wavelengths of LEDs. The inset in each panel shows the constructed LED source with the specific wavelength mentioned in the graph legend.

For instance, to set the illumination period for 25 seconds, the user has to press the "10" button twice and the "1s" button 5 times. Then upon pressing the "OK" button, the light source will stay ON for 25 seconds and then stay OFF for 25 seconds. Until the user presses the "RS" button, the light source will stay ON and OFF for 25 seconds. Fig. 5 shows the illumination period for 1-second, 10-second, and 1-minute time buttons and 25 seconds.

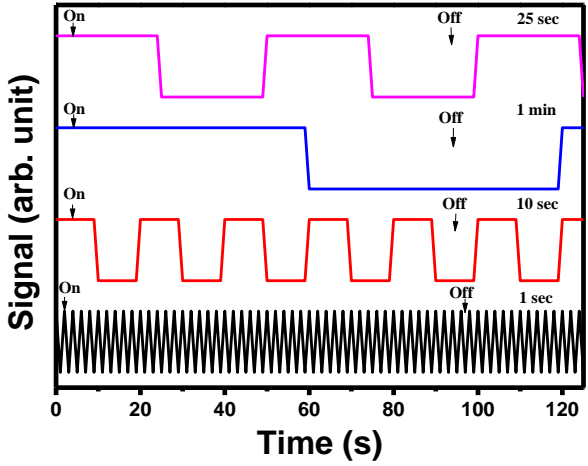

Fig. 5. Illumination period for a different duration.

## IV. PHOTO INDUCED PROPERTIES

As mentioned earlier, the light source could be used where illumination on the sample is required for photophysical or photochemical studies of that particular sample. Here a couple of inorganic solar cells have been used as samples to investigate their photo-induced properties.

### A. Light Wavelength Modulated Impedance Spectroscopy

The light source has been used in investigating light intensity modulated impedance spectroscopy. A solar cell of the structure ITO (~300 nm)/ CdS (~100 nm)/ CdTe (~10 µm) ($CdCl_2$ treated at 390° C for 30 min)/ Cu (~10 nm)/ Ag (~2 µm) was used for this experiment. With the help of an Autolab Potentiostat/Galvanostat-204 from Metrohm, the results shown in Fig. 6 were obtained.

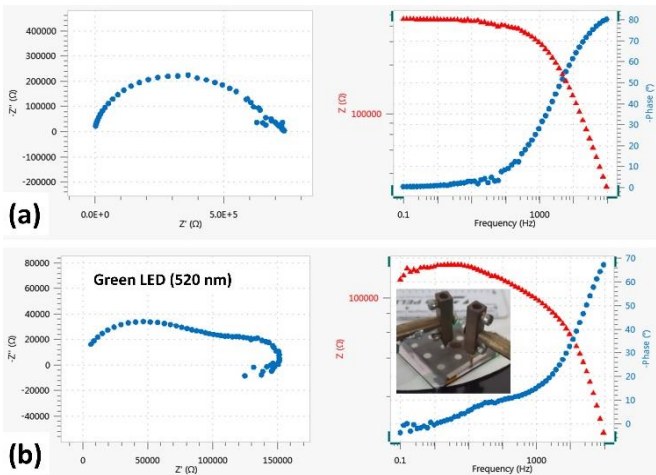

Fig. 6. Impedance Spectroscopy of a CdTe solar cell (a) under dark and (b) under 520 nm LED illumination (CdTe solar cell is shown in the inset).

Fig. 6 (a) shows the Nyquist plot (left) and the Bode plot (right) for the CdTe solar cell without illumination. A semicircle was generated in the Nyquist plot for dark conditions. For the same situation, a high impedance value is also observed at a low frequency in the Bode plot. Figure 6 (b) shows an apparent deviation in shape and magnitude in the Nyquist plot for a LED illumination of 520 nm wavelength (Energy ~2.38 eV). A relatively low value of impedance at low frequency is also evident from the Bode plot. The bandgap energy of CdTe (< 2 eV) [16] matches with the LED illumination energy that introduces these significant changes. Therefore, one can selectively choose LEDs of the desired wavelength to investigate the quality of specific layer materials of any solar cell [17].

### B. Photoresponse of CZTS solar cell with different wavelengths

For finding the photoresponse of a solar cell, a structure of SLG/ FTO/ CdS (~80nm)/ CZTS (Copper Zinc Tin Sulfide) (~1000 nm)/ Ag (2000 nm) has been used. LED wavelengths of 410, 520, and 635 nm were used with 10 second illumination period to determine the photoresponses of this solar cell. Fig. 7 shows the result of this experiment.

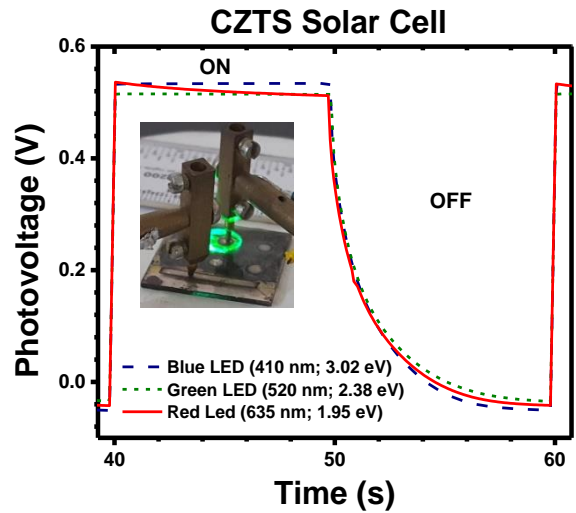

Fig. 7. Photoresponse of a CZTS solar cell for different LED wavelengths (illumination for green LED (520 nm) is shown in the inset).

The CZTS solar cell shows the different responses to different wavelengths of light. We found higher photovoltage for blue light (410 nm) compared to green light (520 nm), as blue light has more energy, generating more energetic electron-hole pairs in the device than green light. On the other hand, red light (635 nm) penetrates deeper into the solar cell and only gets absorbed by the CZTS layer as Red-LED light energy (1.95 eV) is above its' bandgap (<1.6 eV) [18]. Therefore, Red-LED-induced transient photovoltage can assess the quality of the CZTS layer and the transport nature of the photogenerated charge carriers in this layer. A high level of photovoltage was seen initially due to bandgap-matched absorption. As those photoinduced charges have to travel a long path (CZTS layer thickness~1000 nm), they are more likely to recombine within the absorber layer. As a result, we see a significant decline in photovoltage over time for red light (solid curve in Fig.7). In this way, one can selectively characterize different layers of any solar cells.

## V. CONCLUSION

We demonstrated the different functionalities of our LED-based light source by characterizing CdTe and CZTS solar cells, the two most popular inorganic thin film solar cells in the renewable energy research community. We compared the Nyquist and Bode plot of the CdTe solar cell under dark and 520 nm LED illumination to understand the impedance nature of active layers. The transient photovoltage of the CZTS solar cell was studied using three different LEDs to investigate the transport nature of the photogenerated carrier within the CZTS layer within 10 s. The illumination period and level of intensity of LED would be set for a longer duration to study the photodegradation of the same layer. Conventional light sources have limited functionalities to perform this type of study; our constructed light source may overcome those limitations. We anticipate that our portable LED-based light source will also come in handy for other photophysics or photochemistry-based research. The unique features of this light source will pave new ways and possibilities for renewable energy researchers.

ACKNOWLEDGMENT

All the authors gratefully acknowledge the experimental support of the Energy Conversion and Storage Research (ECSR) Section, Industrial Physics Division (IPD), BCSIR Laboratories, Dhaka 1205, Bangladesh Council of Scientific and Industrial Research (BCSIR), under the scope of R&D project # 03-FY2017-2022. The authors also would like to thank Dr. Nipu Kumar Das and Mr. Ashoke Kumar Sen Gupta of the Electrical and Electronic Engineering (EEE), Chittagong University of Engineering and Technology (CUET) for providing solar cell samples during their postgraduate research works at ECSR, IPD, BCSIR. Syed Farid Uddin Farhad also acknowledges the support of TWAS grant # 20-143 RG/PHYS/AS_I for ECSR, IPD.